\documentclass[aps,preprint,showpacs,pra,floatfix]{revtex4-1}
\usepackage{amsmath}
\usepackage{amssymb}
\usepackage[pdftex]{graphicx}
\usepackage{bm}
\usepackage{upgreek}
\usepackage{graphicx}
\usepackage{dcolumn}
\usepackage{bm}
\usepackage{latexsym}

\begin{document}
\title{Possible sorting mechanism for microparticles in an evanescent field}
\author{Eivind Almaas}
\email{Email: eivind.almaas@ntnu.no}
\affiliation{Department of Biotechnology, NTNU - Norwegian University of Science and Technology, N-7491 Trondheim, Norway}
\author{Iver Brevik}
\email{Email: iver.h.brevik@ntnu.no}
\affiliation{Department of Energy and Process Engineering, NTNU - Norwegian University of Science and Technology, N-7491 Trondheim, Norway}

\begin{abstract}
Mie scattering theory is used to calculate radiation forces on a
dielectric microsphere illuminated by evanescent waves, produced by
laser light transmitted obliquely through a flat horizontal dielectric
surface. The incident field is identified with the evanescent field,
and both $p$ and $s$ polarizations are considered. Our investigation
consists of three parts. First, after highlighting the basic
formalism, we report results for the radiation force published in an
earlier paper [J. Opt. Soc. Am. B {\bf 12}, 2429 (1995)], correcting a
few trivial calculational errors.  Second - the main objective of our
paper - is to show how the vertical (lifting) force on microspheres,
typically via a proper adjustment of the laser frequency, can be used
to separate spheres differing by a slight amount in their refractive
index.  This is caused by an oscillatory behavior in the force with
respect to the nondimensional wave number $\alpha$ in the surrounding
medium.  Fine-tuning the wave number $\alpha$, relative to the given
refractive indices in the system, may lead to particle expulsion. The
sorting mechanism turns out to be feasible when $\alpha$ is about
18-20 or larger, which actually is in the region of practical
interest.  Finally, we investigate how variations in the angle of
incidence $\theta_1$ for the laser beam influences the resulting
radiation force.
\end{abstract}


\maketitle

\section{Introduction}
\begin{figure}[t]
\centerline{\includegraphics[width=8cm]{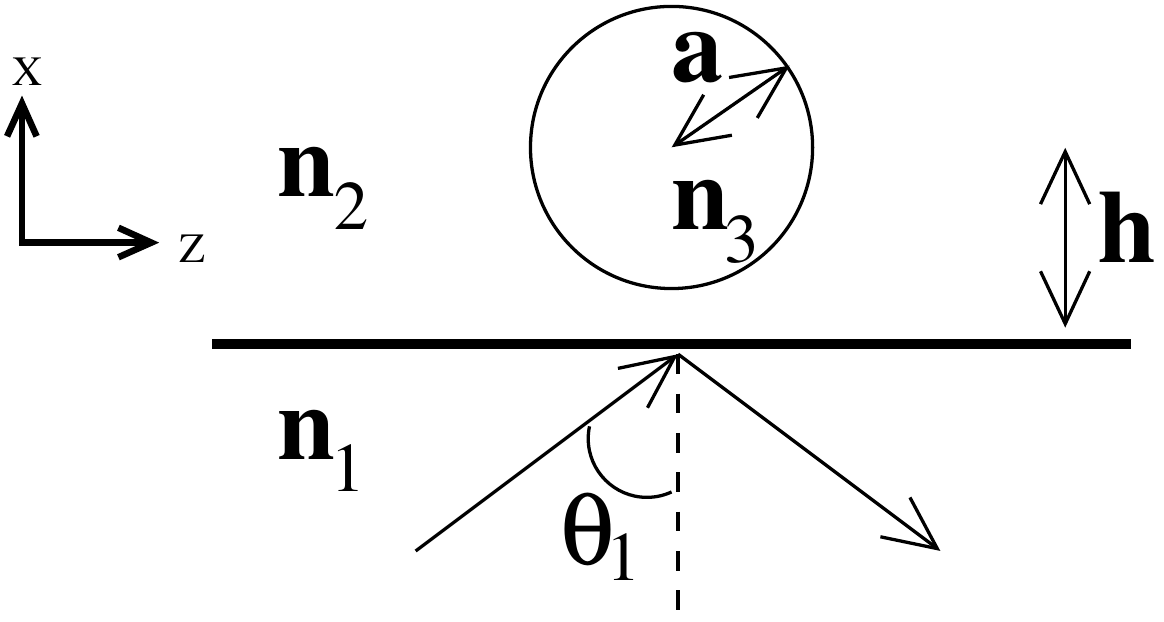}}
\caption{Particle with radius $a$ with distance $h$ between the
  surface and the particle center. The angle of incidence $\theta_1 >
  \theta_{crit}$ for evanescent wave, and indices of refraction $n_1$
  in the lower substrate, $n_2$ in the medium surrounding the sphere,
  and $n_3$ in the sphere.}
\label{fig:geom}
\end{figure}

The guidance and controlled movement of microparticles in an
evanescent field from a laser beam is of considerable theoretical and
practical interest. Theoretically, it is a problem in classical
electromagnetic wave theory.  From a practical point of view, one
would like to construct devices permitting an effective and
non-destructive way of propelling and sorting microparticles, such as
ordinary dielectric particles (e.g. latex spheres), and biological
particles such as red blood cells and bacteria.  Our basic setup is as
sketched in Fig.~\ref{fig:geom} (similar to Ref.~\cite{almaas95}): a
spherical particle of radius $a$ centered at the origin $x=z=0$ is
situated in an evanescent field above a horizontal flat dielectric
surface.  The distance between the plane and the sphere center is
called $h$.  The refractive indices, for simplicity assumed to be
real, are $n_1$ in the lower substrate, $n_2$ in the medium
surrounding the sphere, and $n_3$ in the sphere itself.  A plane laser
beam is incident from below at an angle of incidence $\theta_1$,
greater than the critical angle $\theta_{\rm crit}$ characterizing
total reflection, determined by $\sin \theta_{\rm crit}= n_{21}$ with
$n_{21}=n_2/n_1$.  In experiments it turns out that, when the power
$P$ of the incident laser beam is some hundreds of milliwatts, the
particle is lifted slightly above the surface and is subsequently
moved along the surface at a speed of a few micrometers per
second.  The most common substance for the surrounding medium 2 is
water, with refractive index $n_2=1.33$.  In general, Mie wave theory
is needed in order to describe this situation (cf., for instance,
Ref.~\cite{barton89}), whereas when the nondimensional wave number
$\alpha =2\pi a/\lambda_2$ exceeds about 80 we can make use of the
geometrical optics approximation with sufficient accuracy.

It is probably correct to say that the development of this field began
with the experiment of Kawata and Sugiura in 1992 \cite{kawata92}.
Their experimental setup was as sketched in Fig.~\ref{fig:geom}, and
solid materials, polystyrene latex spheres and glass spheres, were
used.  A Mie theoretical description of the effect was given in
Ref.~\cite{almaas95}, implying the use of an evanescent field taken to
cover the entire incidence region.  Although a limiting factor of this
method is that the electromagnetic boundary conditions at the plate
($x=-h$) are not accounted for, it turned out that this simplified
approach was able to reproduce the experimental observations to a high
level of accuracy.  Later on, several theoretical works have been
published, especially in connection with the trapping of
microparticles in the evanescent field of an optical waveguide with a
step index profile \cite{jaising05}.  An interesting variant is to
consider hollow glass spheres in the evanescent field
\cite{ahluwalia11}.  The theory for absorbing spheres has also been
given \cite{brevik03}.  There are several other related papers, for
instance Ref.~\cite{bekshaev11} studying the internal energy
circulation in light beams, and Ref.~\cite{walz99} dealing with ray
optics calculations for dielectric spheres in an evanescent field.
The review paper \cite{dienerowitz08} is also useful, as are the
dissertations of L{\o}vhaugen \cite{lovhaugen12} and Jaising
\cite{jaising04}.

The purpose of the present paper is twofold: {\bf A.}  We recalculate
and correct some of the expressions for the longitudinal and vertical
radiation force from Ref.~\cite{almaas95}.  A few calculation errors
in the earlier formalism made this undertaking worthwhile.  Since the
formalism is rather complicated, and as the results are of apparent
importance in experimental situations, care should be taken to get
them correct.  Moreover, comparison with similar calculations made
recently by Bekshaev \cite{bekshaev12} makes this recalculation
desirable.  {\bf B.} Our second purpose is to exploit the fact that,
for reasonably large values of the nondimensional wave number, called
$\alpha$, the vertical force on a microsphere in the evanescent field
is an oscillating function of $\alpha$.  For certain narrow
$\alpha$-intervals the vertical force can even be {\it repulsive}.
That means, it is in principle possible to adjust $\alpha$ and other
parameters such that microspheres of given size and given refractive
index are expelled from the main flow of particles with different (and
non-resonating) refractive indices propagating in the evanescent field
above the surface.  We also investigate the non-monotonic dependence
of the force on the angle of incidence.  The possibility of using this
evanescent setup as a sorting device is the key theme of the present
paper.

Let us consider more closely the electromagnetic force acting on a
microsphere.  Assuming a homogeneous interior, the force acts only in
the sphere's boundary layer.  The volume force density is (cf., for
instance, Refs.~\cite{stratton41} or \cite{brevik79})
\begin{equation}
  {\bf f}=-\frac{1}{2}\varepsilon_0 E^2\nabla \varepsilon, \label{1}
\end{equation}
This force should be expected to dominate at the lower end of the
sphere where the evanescent field is strongest.  At first sight this
is somewhat surprising, as one would expect the vertical force to be
{\it attractive}, thus pulling the sphere down, towards the surface.
The situation is however more complicated, at least for the following
three reasons: first, the fields in the interior are concentrated near
the surface, as whispering-gallery modes.  The field power in such
modes are known to be quite large, of the order of hundreds of watts
under usual circumstances (cf., for instance,
Refs.~\cite{rokshari05,brevik10}).  Hence, it is possible that
circulating modes of this sort become totally reflected from the
boundary at the upper part of the sphere, and their interference
giving rise to an outward directed force.  The effects of interference
have been noticed before, for instance by Jaising and Helles{\o}
\cite{jaising05} for evanescent fields near a wave guide.

Second, the influence from electric conductivity $\sigma$ in the
sphere may come into play.  If the surface of the sphere were a
perfect conductor, light rays from below would evidently bounce off
the surface and give rise to a repulsive contribution to the force.
We investigated this point in some detail in Ref.~\cite{brevik03},
together with an analysis of absorptive effects.  Our conclusion was
that a layer of adsorbed film on the sphere's surface, making it
partly conducting, could be an appreciable factor in the observed
repulsive force.  Quantitative estimates for the impurity-induced
conductivity are of course difficult.

Third, one must expect that there are {\it thermophoretic} forces
acting (they are also called photophoretic forces).  By heating one
side of an object, a thermal gradient is established resulting in a
movement away from the hotter region (in our case the maximum
intensity region), towards colder environments. The thermal forces are
known to be strong, up to about 1000 times stronger than radiation
pressure.  Although the relative strength of the thermophoretic force
in the Kawata-Sugiura setup \cite{kawata92} is difficult to estimate,
it seems very probable that the thermal force component is largely
responsible for the observed lifting of the spheres from the surface.
Based upon numerical results in the next section, we derive a lower
threshold for the magnitude of the thermophoretic force in the
Kawata-Sugiura experiment.  Recent treatises on thermophoretic effects
can be found in Refs.~\cite{shvedov10} and \cite{esseling12}.  We
suggest that thermophoretic effects are after all the most important
factor among the three mentioned.

Can this sorting method be used for biological materials?  Probably
not, although the situation is not entirely clear.  For cells, the
situation generally becomes more diffuse since the radii of such
particles are varying.  However, one possibility might be to create a
sorting mechanism also in this case by taking into account the
differences in the refractive index between healthy and sick (or dead)
material.  For living bacteria, many refractive indices are reasonably
well known \cite{ross57}, and the refractive indices are known to be
higher for dead cells than for living cells.  This is related to an
effect which is called {\it dielectrophoresis}.  A cell death is
typically marked by a sharp increase in electric conductivity, as some
ions can more easily pass through newly opened pores in the cellular
membrane.  In the dielectrophoretic analysis, one sorts live cells
from dead ones by arranging for electric field gradients in a narrow
constriction of the cell-carrying fluid.  The inhomogeneous field thus
becomes capable of migrating cells with a conductivity-dependent
velocity.  Recent experimental and theoretical work along these lines
has been presented by Patel {\it et al.} \cite{patel12}.

Consequently, we have to conclude that a sorting method based upon
differences in refractive indices seems most appropriate in cases
where the particle radii are exactly known, as is the case for
monodispersive spheres also called Ugelstad spheres.  For these
situations, it should be possible to separate spheres that differ by a
small amount in their refractive index.  The main experimental
challenge would  be to tune the frequency $\omega$ with a high
degree of accuracy.

In the next section we recapitulate for the sake of readability some
main points of the Mie theory in the form given in
Ref.~\cite{almaas95}, and in Sec. III we present numerical results in
the form of several figures.  Thus Fig.~\ref{fig:corr} corrects some
results from \cite{almaas95}, assuming $n_1=1.75, n_2=1.33$ (water),
and $n_3=\{1.50, 1.60\}$.  Both polarizations $s$ and $p$ are
covered.  A general property inferred from the figure panels is that
the nondimensional vertical forces $Q_x$ when depicted versus the
nondimensional wave number $\alpha$ are negative (attractive), and are
stronger for $p$ polarization than for $s$ polarization. The
longitudinal forces $Q_z$ are positive in all cases thus driving the
microparticles forward, as expected.

Figures~\ref{fig:qx}-\ref{fig:qztheta} show calculated results
pertaining to the proposed sorting method. In order for the setup to
be practically usefule, one key property is that the contrasts between
the refractive indices have to be reasonably large. To keep oversight
over the parameter values, we assume henceforth for the most part that
$n_1$ and $n_3$ have fixed values, $n_1=1.60$ and $n_3=1.50$. Both of
these values are quite standard for dielectrics. The most important
remaining parameter is thus $n_2$.

One may ask: is the sorting method useful for {\it liquids}?  The
answer seems to be no, as the requirement about contrast is not
fulfilled. There are some liquids that are known to have low
refractive index (for instance the liquid called fluorine refrigerant
R-22 has $n=1.26$), but even in such a case the contrast turns out to
be insufficient. The conclusion is that one has most likely to resort
to the case of a {\it gas} as ambient medium 2. For gases, the
refractive indices are very close to unity; even for the extreme case
of benzene the value of $n$ is only 1.00176. So, in the following we
assume that $n_2$ is equal to unity, or close to it.  As the figures
will show, if $\alpha$ is adjusted accurately enough, it is in
principle possible to expel selected microparticles from the main flow
traveling in the evanescent field above a planar surface.

\section{Extracts of the basic formalism}

Here, we present the basics of the formalism (for more details,
cf. Refs.~\cite{almaas95}, \cite{barton89}, and \cite{bekshaev12A}).
Let ${\bf E}^i$ and ${\bf H}^i$ denote the fields incident on the
sphere in medium 2.  In the following, we shall only need the radial
components $E_r^i$ and $H_r^i$, and the nondimensional wave number of
the incident field is
\begin{equation}
\alpha= k_2a=n_2\omega a/c. \label{2}
\end{equation}
The radial part of the Helmholtz equation allows us to expand the
fields as
\begin{equation}
E_r^i=\frac{E_0}{\tilde{r}^2}\sum_{l=1}^\infty \sum_{m=-l}^l l(l+1)A_{lm}\psi_l(\alpha \tilde r)Y_{lm}(\Omega), \label{3}
\end{equation}
\begin{equation}
H_r^i=\frac{H_0}{\tilde{r}^2}\sum_{l=1}^\infty \sum_{m=-l}^l l(l+1)B_{lm}\psi_l(\alpha \tilde r)Y_{lm}(\Omega), \label{4}
\end{equation}
where $E_0$ and $H_0$ are field amplitudes, related to each other via
$H_0=\sqrt{\epsilon_0/\mu_0}E_0$.  Moreover $\tilde{r}=r/a$ is the
nondimensional radius, and
\begin{equation}
\psi_l(x)=xj_l(x)=\sqrt{\frac{\pi x}{2}}\,J_\nu (x) \label{5}
\end{equation}
with $\nu=l+1/2$ is the Riccati-Bessel function.  The spherical
harmonic is $Y_{lm}(\Omega)$ with $\Omega=(\theta, \phi)$, $\theta$
and $\phi$ being respectively the polar and the azimuthal angles, and
the time factor $e^{-i\omega t}$ has been omitted.

When $E_r^i$ and $H_r^i$ are known, the coefficients $A_{lm}$ and
$B_{lm}$ can be found as
\begin{equation}
A_{lm}=\frac{1}{E_0l(l+1)\psi_l(\alpha)}\int_\Omega E_r^i(a, \theta, \phi)Y_{lm}^*(\Omega)d\Omega, \label{6}
\end{equation}
\begin{equation}
B_{lm}=\frac{1}{H_0l(l+1)\psi_l(\alpha)}\int_\Omega H_r^i(a, \theta, \phi)Y_{lm}^*(\Omega)d\Omega, \label{7}
\end{equation}
with $d\Omega=\sin\theta d\theta d\phi$.  Here the integration is
taken over the whole spherical surface (in principle, the surface has
an arbitrary radius set equal to $r=b$ in Ref.~\cite{almaas95}, but we
simplify the formalism by putting $b=a$).

We now identify the incident field ${\bf E}^i $ with the evanescent
field. Note that at this point, the presence of the substrate between
media 1 and 2 is ignored; we let the mathematical expression for the
evanescent field be extended to all negative values for the vertical
coordinate $x$.  Let the origin $x=y=z=0$ be placed in the center of
the sphere, which again lies at a height $h$ above the substrate. Let
$T_\parallel$ and $T_\perp$ denote the transmission coefficients for
the field lying respectively in the plane of incidence ($p$
polarization) and normal to it ($s$ polarization),
\begin{equation}
T_\parallel =\frac{E_\parallel^{(2)}}{E_\parallel^{(1)}}=
\frac{2n_{21}\cos \theta_1}{n_{21}^2\cos \theta_1+i(\sin^2\theta_1-n_{21}^2)^{1/2}}, \label{8}
\end{equation}
\begin{equation}
T_\perp = \frac{E_\perp^{(2)}}{E_\perp^{(1)}} =\frac{2\cos \theta_1}{\cos \theta_1+i(\sin^2\theta_1-n_{21}^2)^{1/2}}. \label{9}
\end{equation}
Here $\theta_1$ is the angle of incidence in medium 1, and
$n_{21}=n_2/n_1$. With the abbreviations
\begin{equation}
\beta=\frac{n_1\omega}{c}(\sin^2\theta_1-n_{21}^2)^{1/2}, \quad \gamma=\frac{n_1\omega}{c}\sin \theta_1 \label{10}
\end{equation}
we can then express the radial component of the incident field as
\[ E_r^i=\Big\{ \frac{1}{n_{21}}T_\parallel E_\parallel^{(1)}[ \sin \theta_1\sin\theta \cos \phi -i(\sin^2\theta_1-n_{21}^2)^{1/2}\cos\theta]  \]
\begin{equation}
+T_\perp^{(1)}\sin \theta \sin\phi \Big\}  \exp[-\beta (x+h)+i\gamma z]. \label{11}
\end{equation}
Together with an analogous expression for the magnetic field we can
now calculate the coefficients $A_{lm}$ and $B_{lm}$ from
Eqs.~(\ref{6}) and (\ref{7}).  We do not go into further detail here,
but mention the following useful relations between the $s$ and $p$
polarizations,
\begin{equation}
A_{lm}(s-{\rm pol.})=\frac{T_\perp}{n_2T_\parallel}B_{lm}(p-\rm pol.), \label{12}
\end{equation}
\begin{equation}
B_{lm}(s-{\rm pol.})=-\frac{n_2T_\perp}{T_\parallel}A_{lm}(p-\rm pol.). \label{13}
\end{equation}
In Ref.~\cite{almaas95} we expressed $A_{lm}$ and $B_{lm}$ in terms of
three integral quantities called $Q_1, Q_2$, and $Q_3$.  Here, the
expression (46) for $Q_2$ should have been multiplied with a factor 2,
as well as the expression (49) for $Q_3$. Combined with some minor
errors in the numerical code, a complete recalculation is desirable.
We give the correct results in the next section.

\section{Calculated results, and discussion}

\begin{figure}[t]
\centerline{\includegraphics[width=16cm]{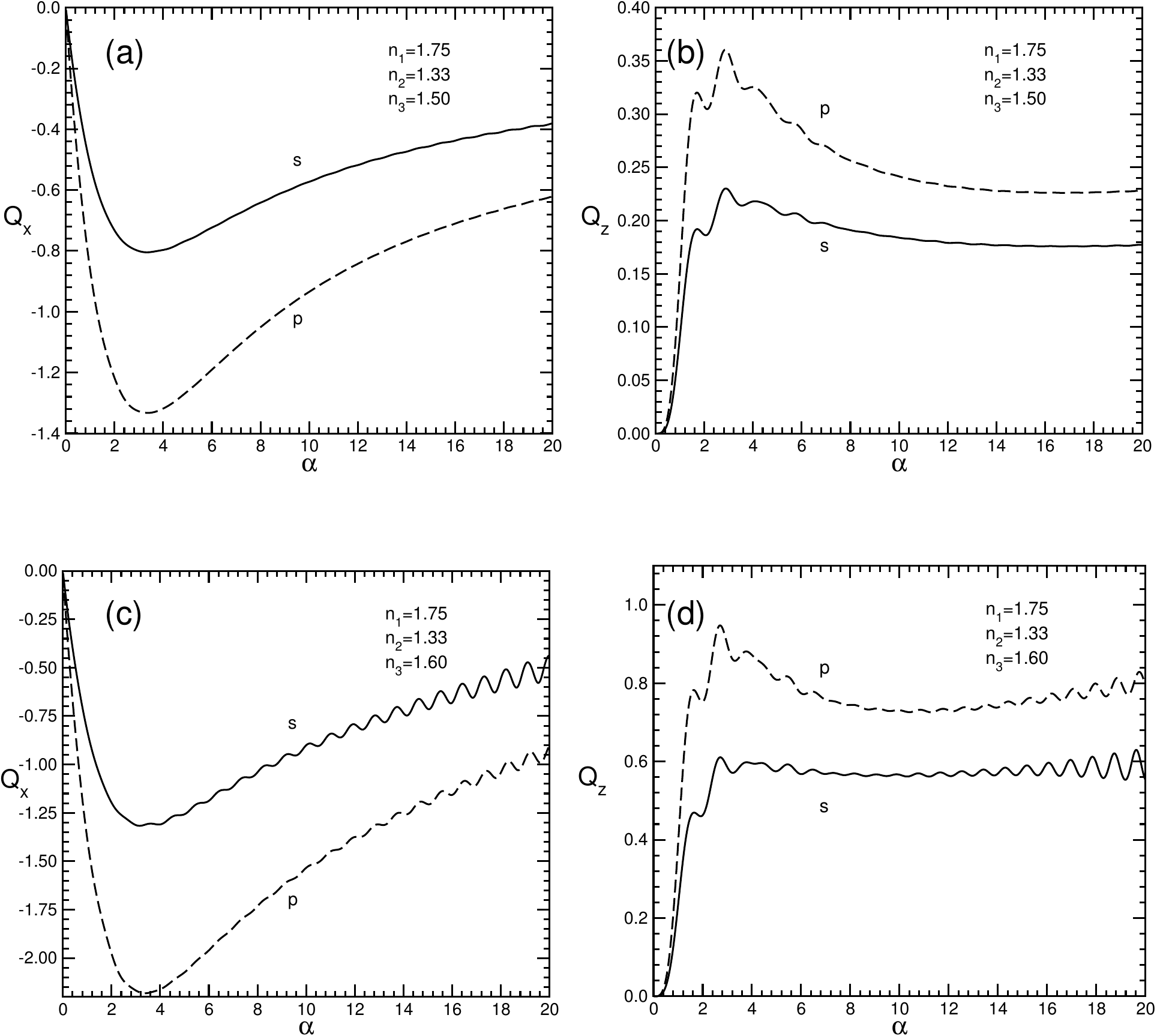}}
\caption{Corrected force calculations from
  Ref. \cite{almaas95}. Panels (a)-(d) correspond to Figs. 4, 5, 6 and
  7 in \onlinecite{almaas95}. Here, $Q_x$ ($Q_z$) is the
  nondimensional vertical (horizontal) radiation force as defined in
  Eq. (\ref{15}). Notation $s$ and $p$ indicates the incident wave
  polarization.}
\label{fig:corr}
\end{figure}

Following the notation of Ref.~\cite{almaas95}, we let $F_x$ denote
the vertical and $F_z$ the horizontal force on the sphere. Their
nondimensional counterparts are defined as
\begin {equation}
Q_x=\frac{F_x}{\varepsilon_0E_0^2a^2}, \quad Q_z=\frac{F_z}{\varepsilon_0E_0^2a^2}. \label{14}
\end{equation}
Figure~\ref{fig:corr}, panels (a)-(d), replace figures 4-7 in our
earlier article \cite{almaas95}.  In contrast to the original article,
where the nondimensional wave number $\alpha \le 10$ was moderate, we
have now been able to explore a much larger range of $\alpha$.  We may
now clearly see the oscillatory behavior which is typical for large
$\alpha$, even for this choice of refractive index in medium 2,
$n_2=1.33$ (panels (c) and (d)).  Our definition for the polarizations
$p$ and $s$ are as usual: $p$ polarization is when the field component
$E_\parallel^{(1)}$ in the substrate (medium 1) lies in the plane of
incidence, whereas $s$ polarization is when the corresponding
component $E_\perp^{(1)}$ is orthogonal to this plane.

From the panels in Fig.~\ref{fig:corr} it is seen, as mentioned above,
that the absolute magnitude of the vertical force $Q_x$ is biggest for
the case of $p$ polarization. The same is true for the horizontal
force $Q_z$.  Of main interest here is evidently $Q_x$.  This force is
negative, as expected in this range of $\alpha$.  It means that the
radiation force seeks to pull the sphere down towards the surface.
The force $Q_z$ pushes the sphere forward, as mentioned.  Typical
horizontal velocities in the Kawata-Sugiura experiment, with water
surroundings (medium 2), were in the region $v \sim 1-2~\mu$m/s.  Note
that the figures refer to the case where the sphere rests upon the
surface, i.e., $h=a$. The differences between panels (a)-(d) and the
figures 4-7 in Ref.~\cite{almaas95} are seen to be rather significant.

As a by-product of these figures, we can use them to estimate the
lower threshold for the lifting force observed in the Kawata-Sugiura
experiment \cite{kawata92}.  According to these authors, "the particle
is forced to float from the substrate surface and to slide along the
surface".  That means, there must be a lifting force which we shall
call $F_{\rm lift}$, attributed to thermophoresis in the previous
section, strong enough to overcome gravity together with the radiation
force. It must thus satisfy the inequality
\begin{equation}
F_{\rm lift} > mg +|F_x|. \label{15}
\end{equation}
It is instructive to consider a concrete example (cf. also
Ref.~\cite{brevik03}).  Let the incident laser beam power in vacuum be
$P=150~$mW, distributed over a circular cross-sectional area of
diameter 10 $\mu$m.  Then the Poynting vector becomes
$(\varepsilon_0/2)cE_0^2=19.0$ MW/m$^2$=19.0 $\mu$W/$\mu$m$^2$.
Taking the radius of the sphere to be $a=1$ $\mu$m, we calculate
$\varepsilon_0 E_0^2a^2=0.13$ pN.  If the density of the sphere is 2.4
g/cm$^3$ (glass), the weight of it becomes 0.10 pN.  Assuming Nd:YAG
laser light with fundamental wavelength 1.06 $\mu$m in vacuum, the
wavelength in the surrounding medium 2 (assumed to be water with
$n_2=1.33$) becomes $\lambda_2=0.80$ $\mu$m, giving $\alpha=2\pi
a/\lambda_2=7.9$.  From Fig.~\ref{fig:corr}(a) we read off
$Q_x=-1.05$, in the case of $p$ polarization.  From Eq.~(\ref{15}) we
thus get
\begin{equation}
F_{\rm lift} > mg+(\varepsilon_0E_0^2a^2)|Q_x|=(0.10+0.13 \times 1.05){\rm pN}=0.24~\rm pN. \label{16}
\end{equation}
The sphere's weight, and the vertical radiation force, are thus in this case comparable.

\begin{figure}[t]
\centerline{
\hspace*{-1cm}
\includegraphics[width=10cm]{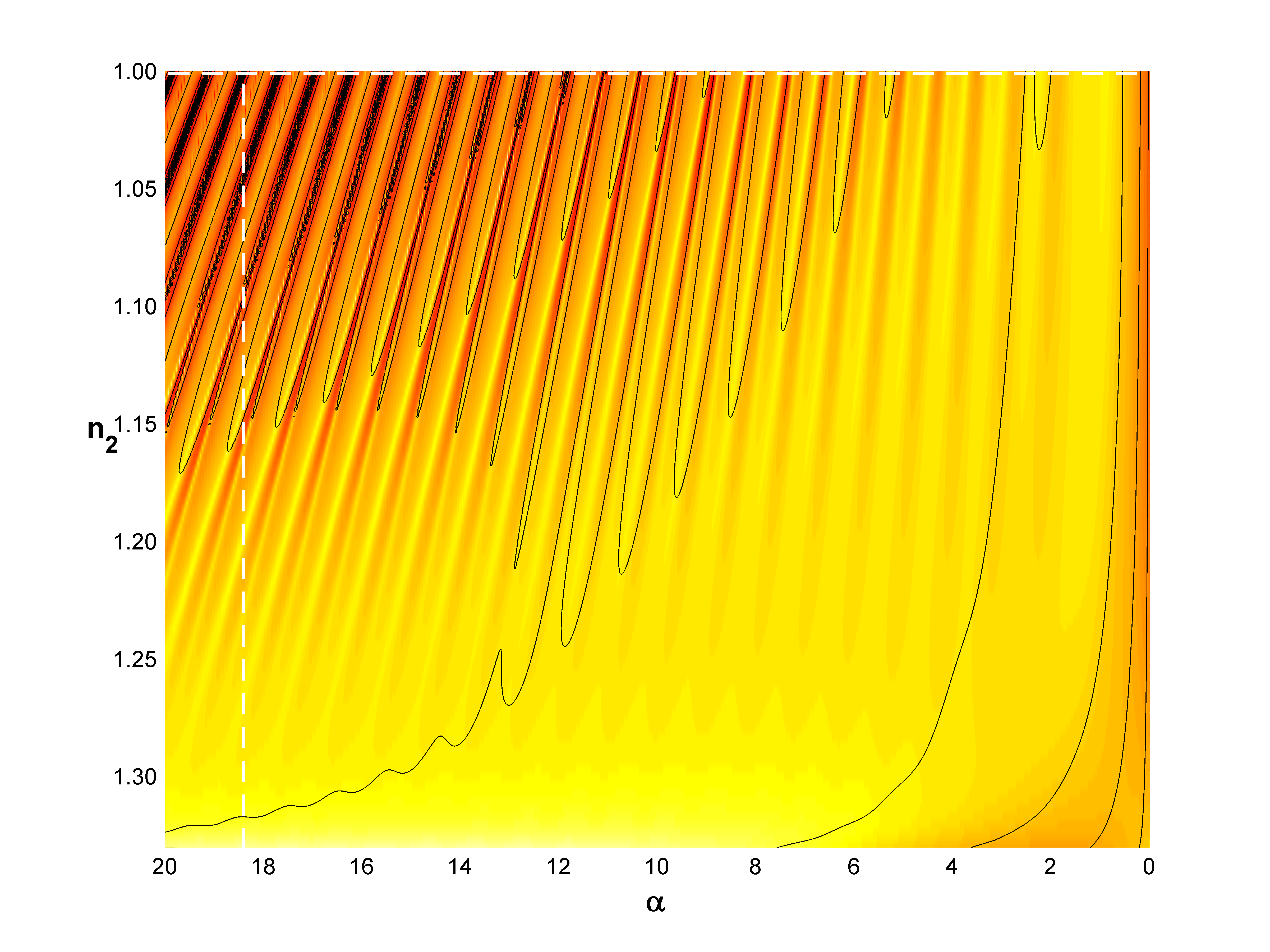}
\hspace*{-1cm}
\includegraphics[width=8cm]{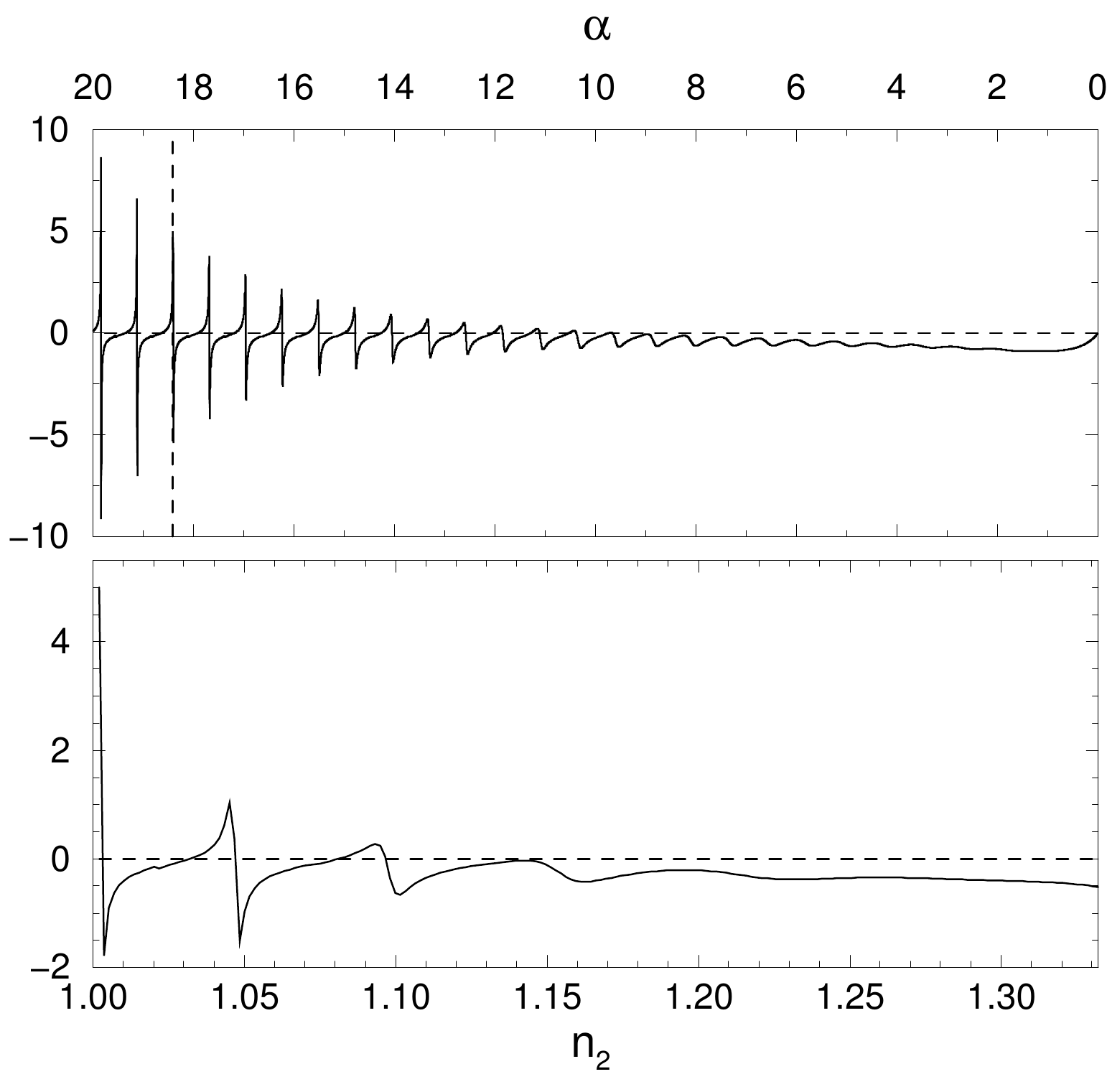}
}
\setlength{\unitlength}{1cm}
\begin{picture}(1,1)
        \put(-7.5,8.2){\Large\bf (a)}
        \put(7.5,7.3){\Large\bf (b)}
        \put(7.5,4){\Large\bf (c)}
\end{picture}
\caption{(Color online) {\bf (a)} Nondimensional vertical force
  $Q_x(\alpha,n_2)$ for $s$-polarized incident beam as function of
  particle size parameter $\alpha$ and the index of refraction in
  medium 2, $n_2$. Other parameters are the angle of incidence
  $\theta_1=57^\circ$, and the refractive indices $n_1=1.60$, and
  $n_3=1.50$. Panels {\bf (b)} and {\bf (c)} display $Q_x(\alpha,n_2)$
  along the (white) dashed lines in (a); $Q_x(\alpha,1.00)$ and
  $Q_x(18.406,n_2)$, respectively. The dashed lines in (b) and (c) are
  guides to the eye.}
\label{fig:qx}
\end{figure}

\begin{figure}[t]
\centerline{
\hspace*{-1cm}
\includegraphics[width=10cm]{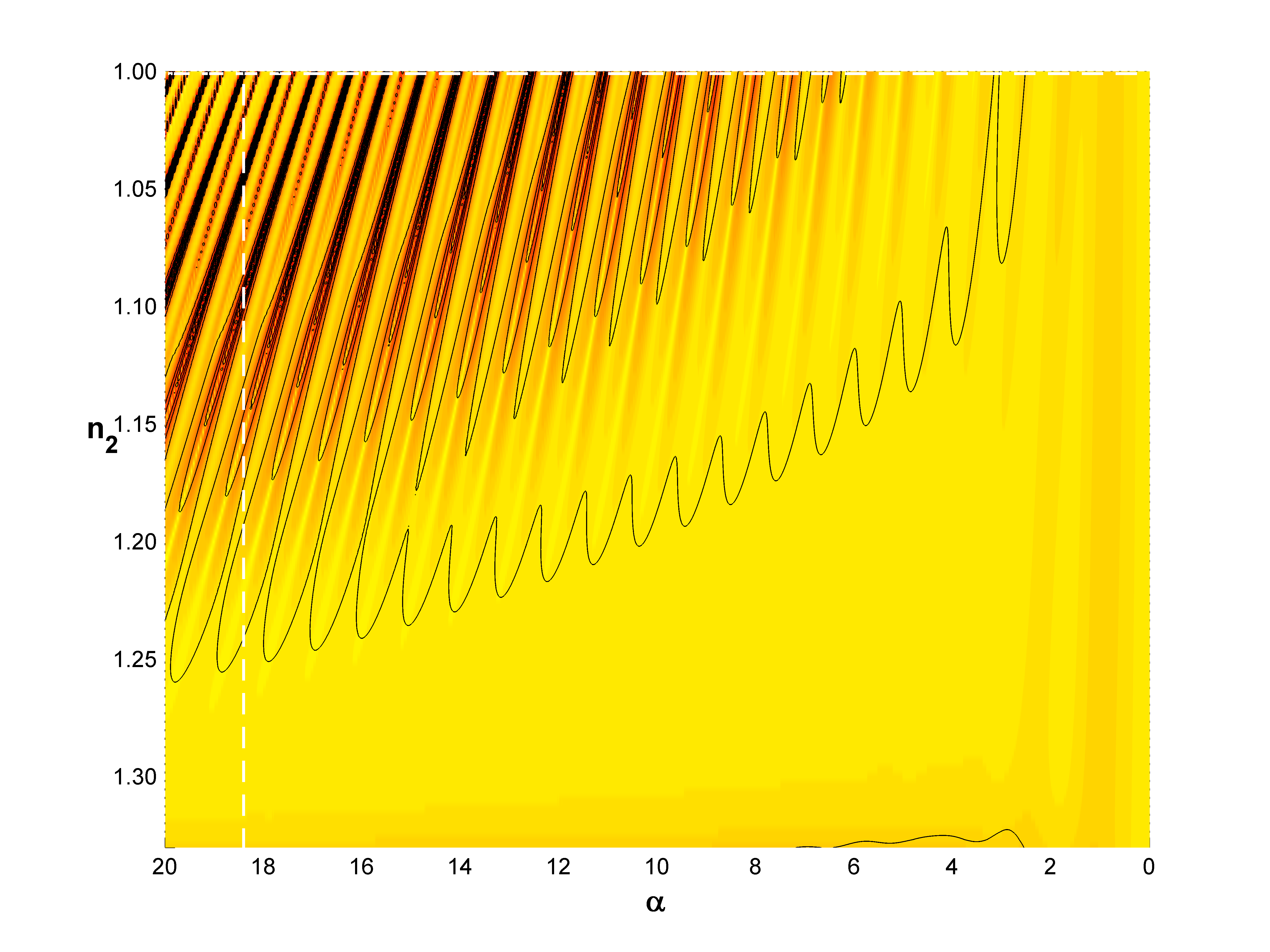}
\hspace*{-1cm}
\includegraphics[width=8cm]{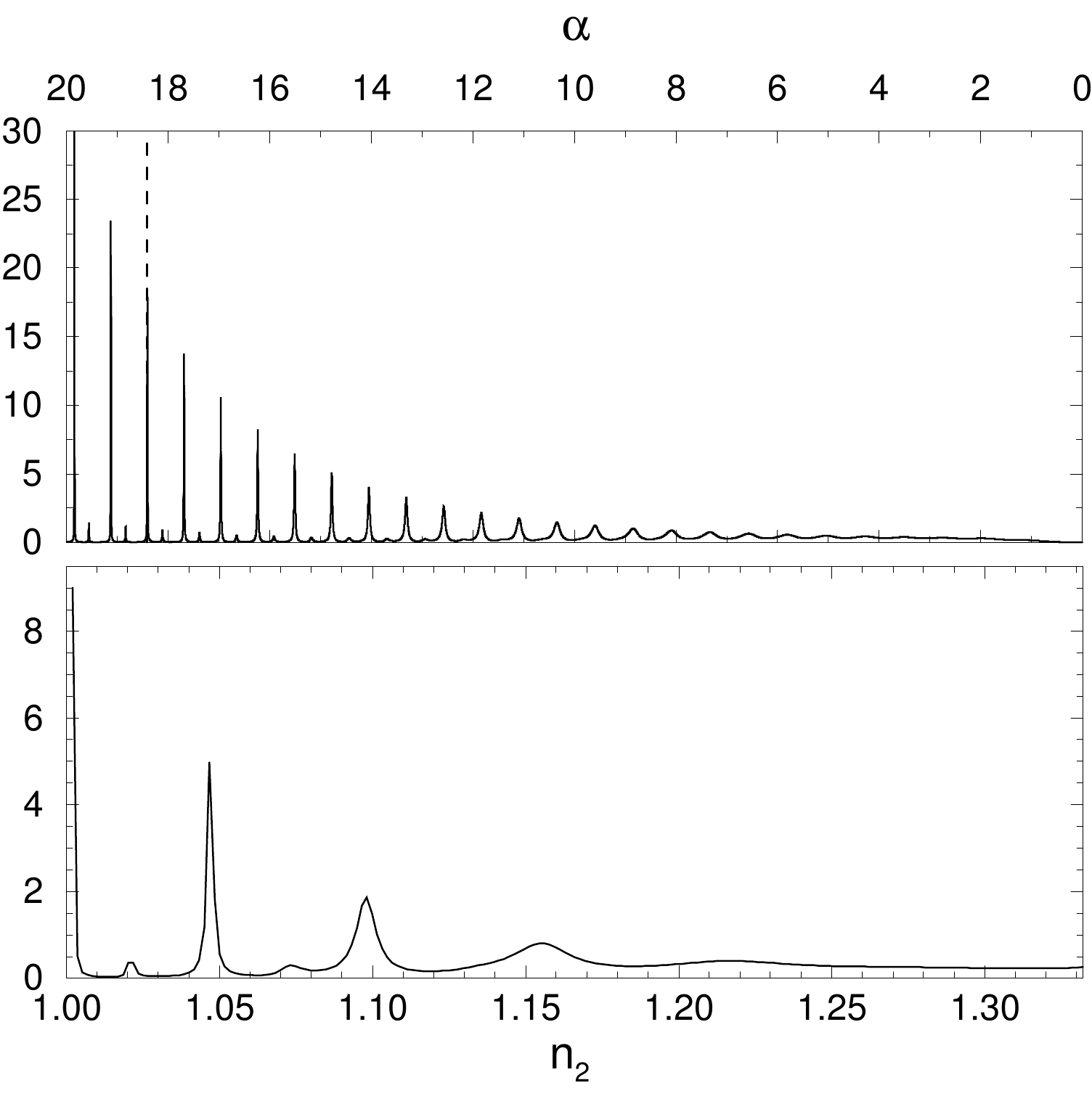}
}
\setlength{\unitlength}{1cm}
\begin{picture}(1,1)
       \put(-7.5,8.2){\Large\bf (a)}
       \put(7.5,7.3){\Large\bf (b)}
       \put(7.5,4){\Large\bf (c)}
\end{picture}
\caption{(Color online) {\bf (a)} Nondimensional horizontal force
  $Q_z(\alpha,n_2)$ for $s$-polarized incident plane wave as function
  of particle size parameter $\alpha$ and the index of refraction in
  medium 2, $n_2$. Other parameters are the angle of incidence
  $\theta_1=57^\circ$, and the refractive indices $n_1=1.60$, and
  $n_3=1.50$. Panels {\bf (b)} and {\bf (c)} display $Q_x(\alpha,n_2)$
  along the (white) dashed lines in (a); $Q_x(\alpha,1.00)$ and
  $Q_x(18.406,n_2)$, respectively. The dashed lines in (b) and (c) are
  guides to the eye.}
\label{fig:qz}
\end{figure}

\begin{figure}[t]
\centerline{
\includegraphics[width=14cm]{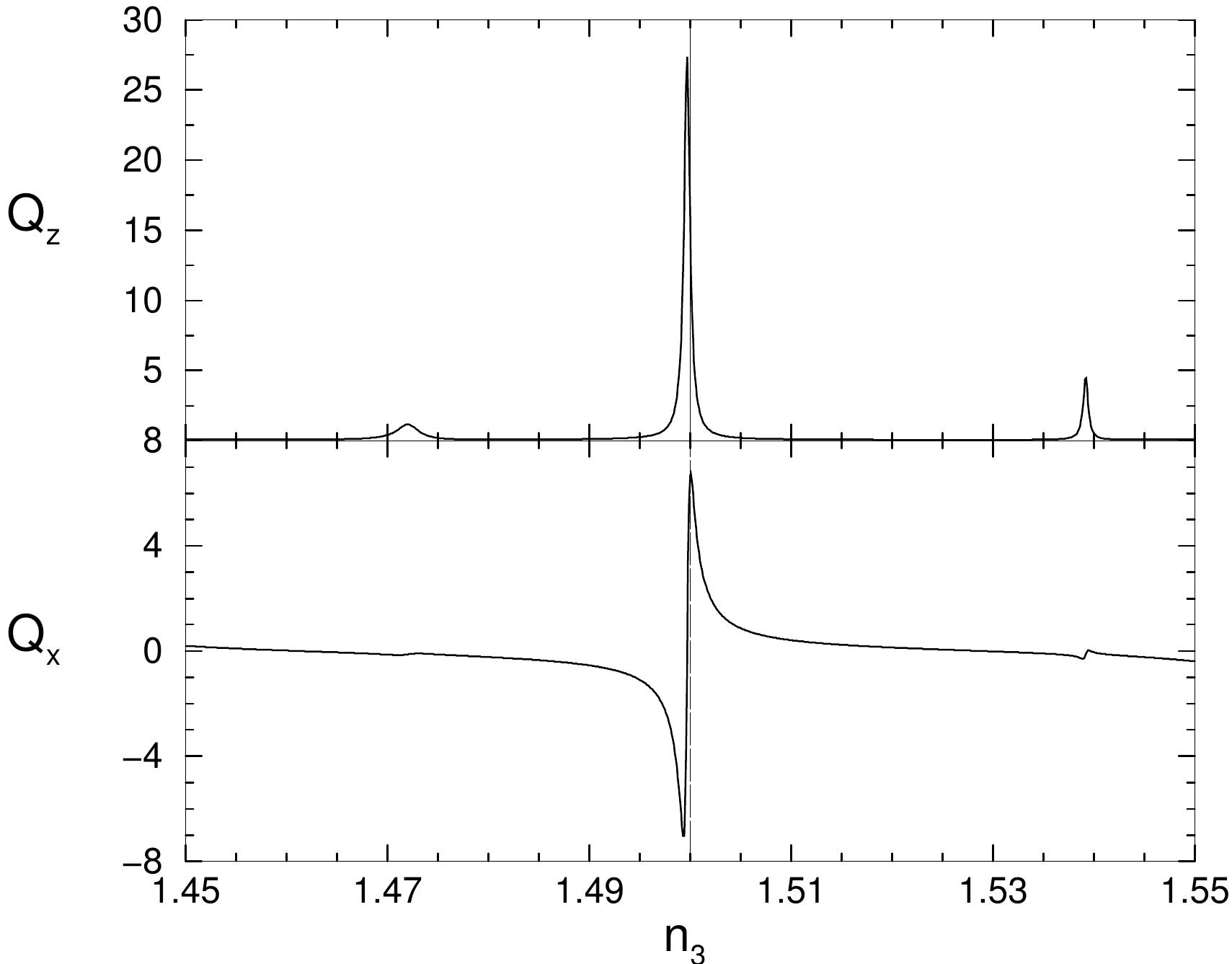}
}
\caption{Nondimensional vertical $Q_x$ and horizontal $Q_z$ force as
  function of refraction index in microsphere $n_3$, for $\alpha =
  18.406$, $\theta_1 = 51^\circ$, $n_1=1.6$, and $n_2=1.0$. The dashed
  vertical line at $n_3=1.50$ is a guide to the eye.}
\label{fig:rep}
\end{figure}

\begin{figure}[t]
\centerline{
\hspace*{-1cm}
\includegraphics[width=10cm]{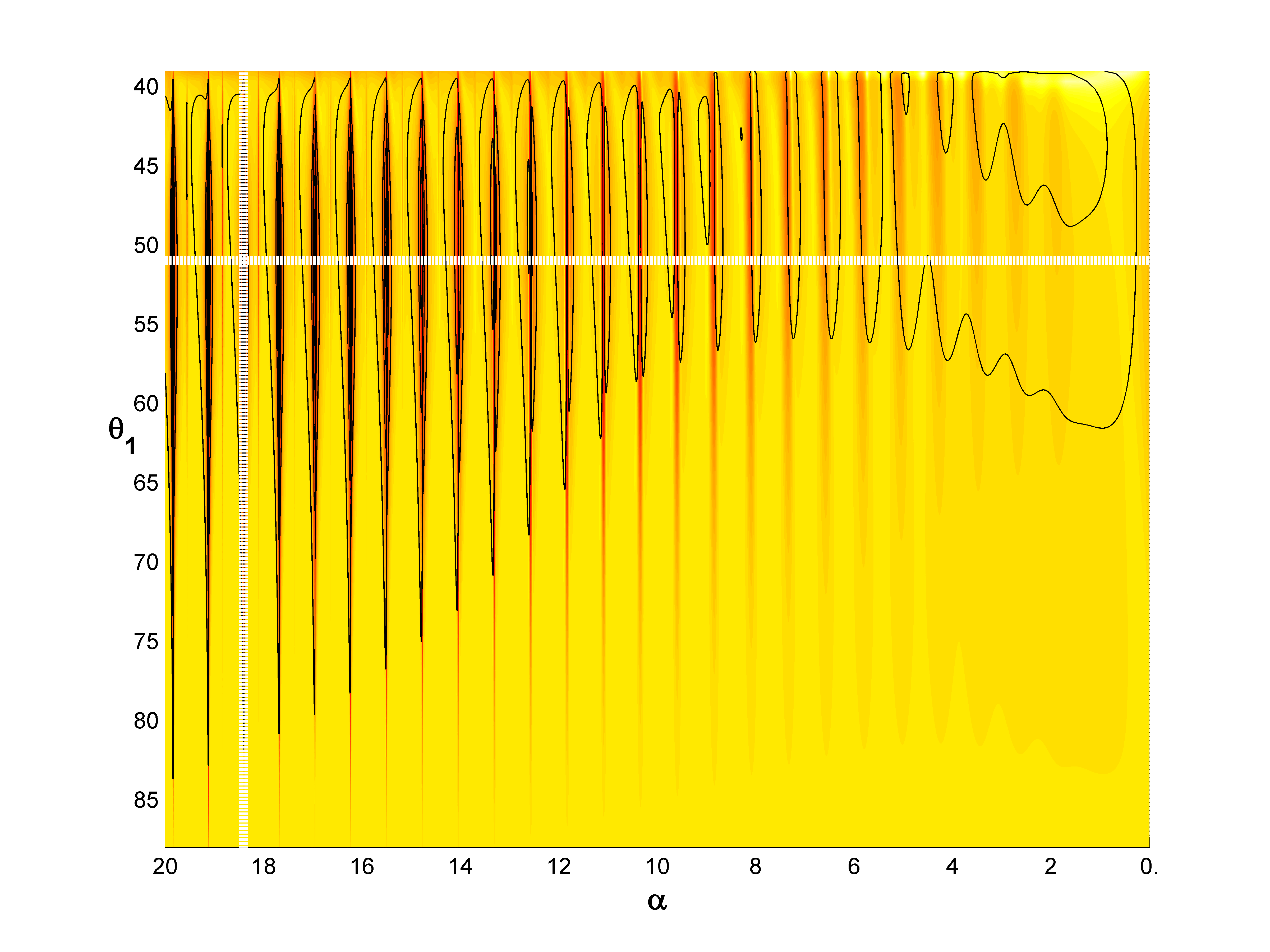}
\hspace*{-1cm}
\includegraphics[width=8cm]{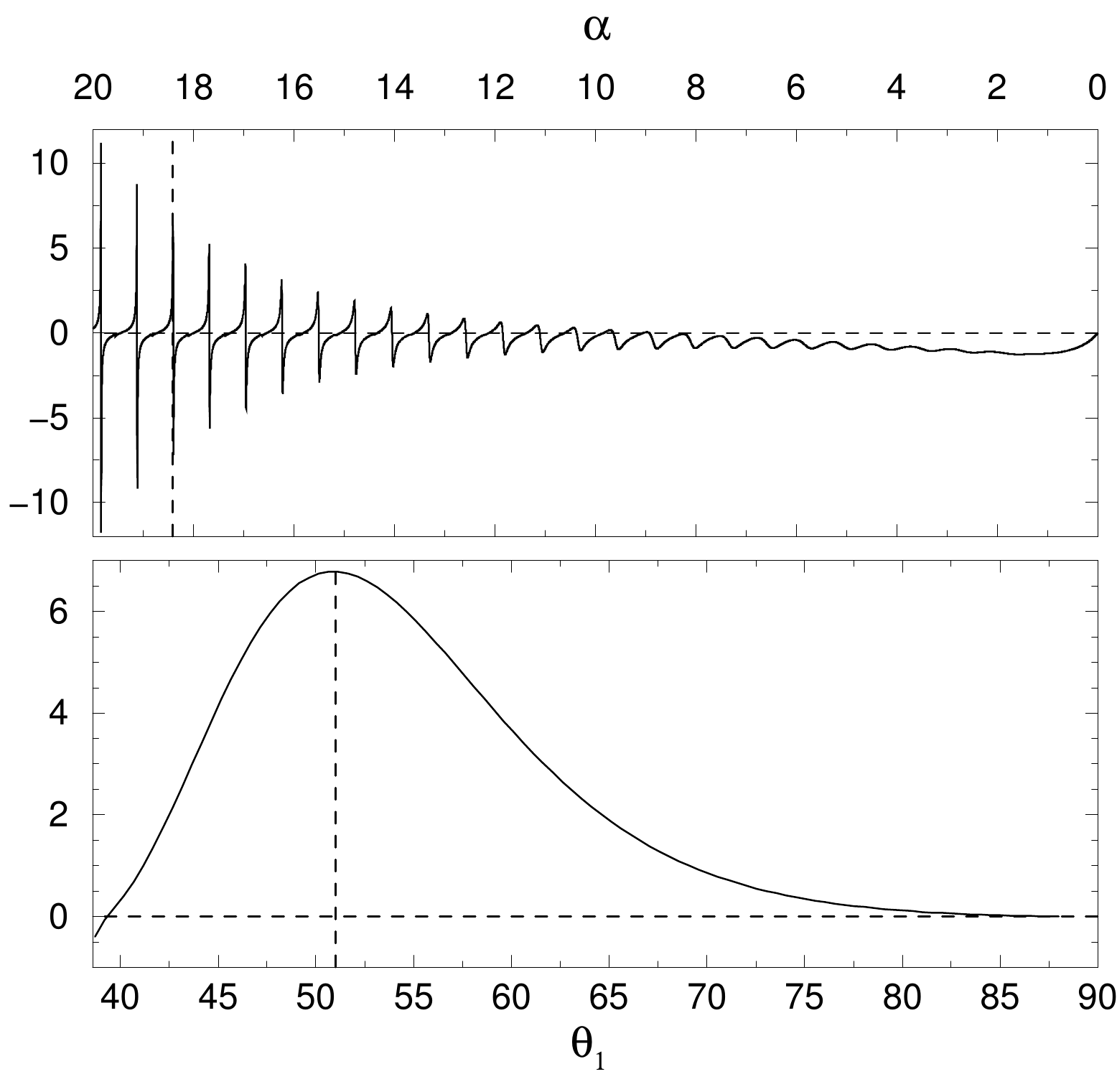}
}
\setlength{\unitlength}{1cm}
\begin{picture}(1,1)
        \put(-7.5,8.2){\Large\bf (a)}
       \put(7.5,7.3){\Large\bf (b)}
       \put(7.5,4){\Large\bf (c)}
\end{picture}
\caption{(Color online) {\bf (a)} Nondimensional vertical force
  $Q_x(\alpha,\theta_1)$ for $s$-polarized incident plane wave as
  function of particle size parameter $\alpha$ and angle of incidence
  $\theta_1>\theta_{crit}$. The refractive indicies are $n_1=1.60$,
  $n_2=1.00$, and $n_3=1.50$. Panels {\bf (b)} and {\bf (c)} display
  $Q_x(\alpha,\theta_1)$ along the (white) dashed lines in (a);
  $Q_x(\alpha,51^\circ)$ and $Q_x(18.406,\theta_1)$, respectively. The
  dashed lines in (b) and (c) are guides to the eye. }
\label{fig:qxtheta}
\end{figure}

\begin{figure}[t]
\centerline{
\hspace*{-1cm}
\includegraphics[width=10cm]{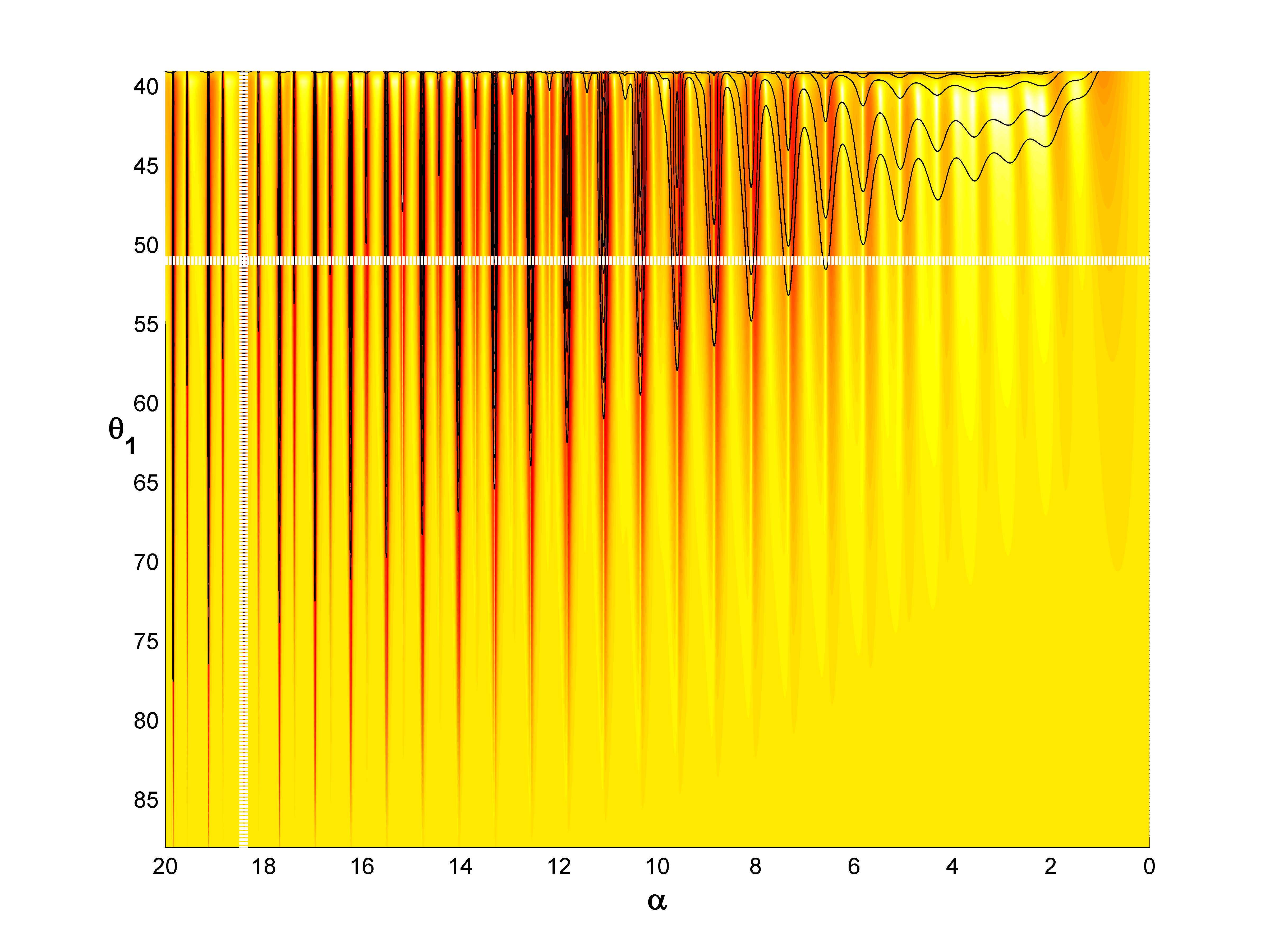}
\hspace*{-1cm}
\includegraphics[width=8cm]{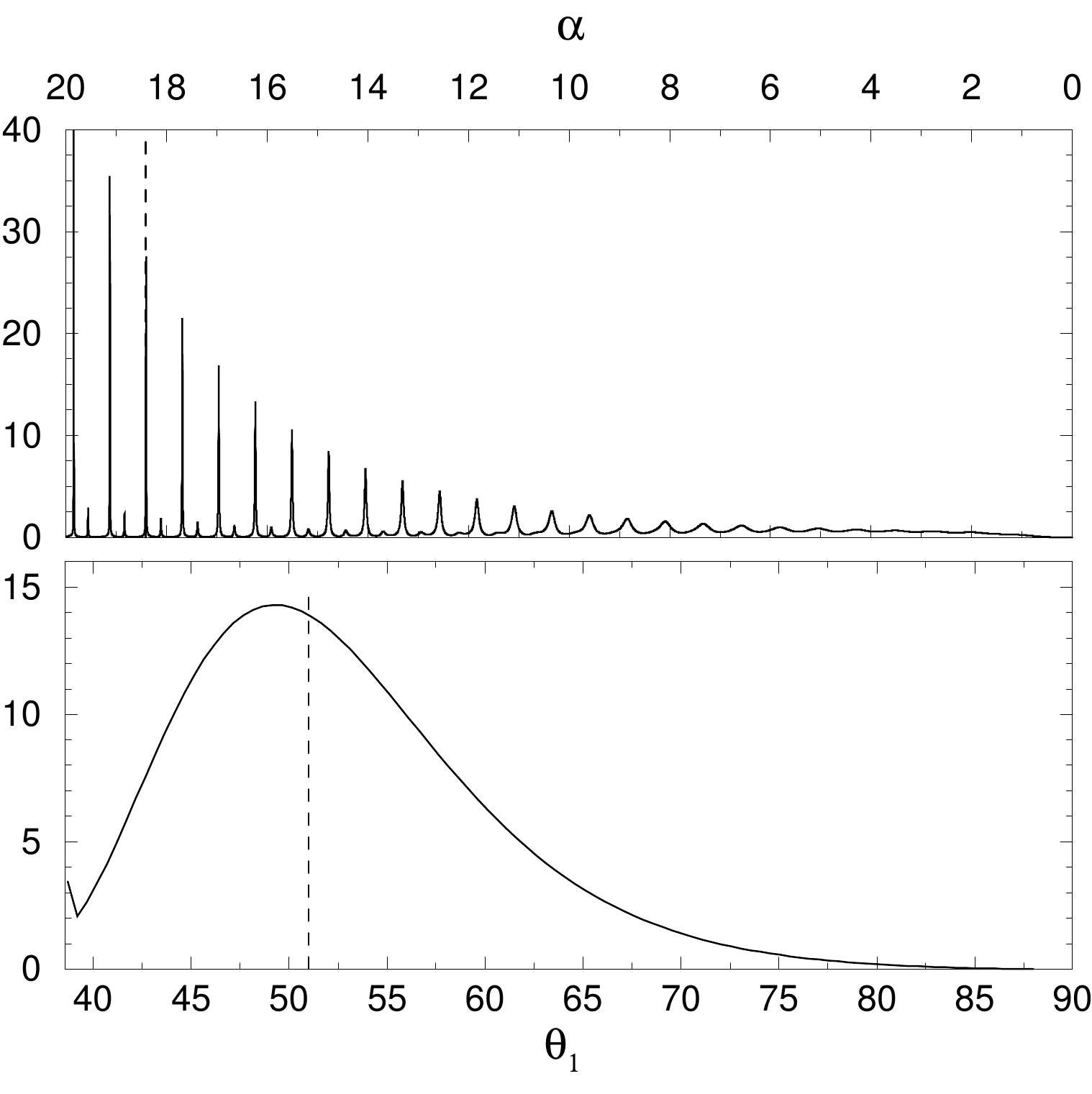}
}
\setlength{\unitlength}{1cm}
\begin{picture}(1,1)
        \put(-7.5,8.2){\Large\bf (a)}
       \put(7.5,7.3){\Large\bf (b)}
       \put(7.5,4){\Large\bf (c)}
\end{picture}
\caption{(Color online) {\bf (a)} Nondimensional horizontal force
  $Q_z(\alpha,\theta_1)$ for $s$-polarized incident plane wave as
  function of particle size parameter $\alpha$ and angle of incidence
  $\theta_1>\theta_{crit}$. The refractive indicies are $n_1=1.60$,
  $n_2=1.00$, and $n_3=1.50$. Panels {\bf (b)} and {\bf (c)} display
  $Q_z(\alpha,\theta_1)$ along the (white) dashed lines in (a);
  $Q_z(\alpha,51^\circ)$ and $Q_x(18.406,\theta_1)$, respectively. The
  dashed lines in (b) and (c) are guides to the eye. }
\label{fig:qztheta}
\end{figure}

Figures~\ref{fig:qx}-\ref{fig:qztheta} present our new
results. Figure~\ref{fig:qx} is probably the one of main interest, as
for a reasonable range of parameters $\{ \alpha, n_2\}$ it shows how
the vertical force $Q_x$ may expel selected microparticles from the
main flow in the evanescent field above the planar surface. This
requires, of course, a positive value of $Q_x$. Panel~\ref{fig:qx}(b)
corresponds to $n_2=1$ (gas). If $\alpha$ is about 18 ($a/\lambda_2$
is about 3), which is a reasonable value for microparticles, we see
that $Q_x$ can reach a large value of about 5.  This should be quite
sufficient to give the selected particles a significant outward
kick. From panel~\ref{fig:qx}(c) it is seen that also for a larger
value of $n_2$ (around 1.05), there is the possibility to obtain a
significant outward force. Here, we have chosen the fixed wave number
value $\alpha=18.406$, since it corresponds to a local force maximum
(see panel (b)).

Singular behaviors of the same kind are also found for the horizontal
force $Q_z$, as shown in Fig.~\ref{fig:qz}. This is as one should
expect. The horizontal force is however of secondary importance in the
present problem.  In addition to treating $n_2$ as the only adjustable
parameter, it is of interest to investigate how different values of
the refractive index $n_3$ in the sphere influence the force. This is
illustrated in Fig.~\ref{fig:rep}, assuming gas surroundings ($n_2=1$)
and the fixed nondimensional wave number $\alpha=18.406$. We observe the
presence of a sudden switch in the sign of $Q_x$ when $n_3$ is slightly
less than 1.50.

Finally, we have calculated the effect of using different values of
the angle of incidence $\theta_1$. Panel~\ref{fig:qxtheta}(a) shows
how the vertical force varies versus values of the parameter set $\{
\alpha, \theta_1\}$. Panel~\ref{fig:qxtheta}(b) demonstrates the sharp
peaks versus $\alpha$ when $\theta_1$ is kept fixed ($51^\circ$), and
panel~\ref{fig:qxtheta}(c) shows how the vertical force varies with
$\theta_1$ when $\alpha$ is kept constant (18.406).
Figure~\ref{fig:qztheta} shows analogous results for the horizontal
force $Q_z$.

To conclude: our theoretical investigations indicate that a sorting
mechanism for selected microparticles in the evanescent field may
under certain conditions be feasible. To investigate whether the
method is useful in a practical application, one has to proceed to
experimental tests.

\bigskip

{\bf Acknowledgments} We thank Olav Gaute Helles{\o} and P{\aa}l
L{\o}vhaugen in Troms{\o}, and Aleksandr Bekshaev in Odessa, for
valuable discussions and correspondence.

\newpage

\newpage

\newpage

\newpage

\newpage

\newpage

\newpage

\end{document}